\documentclass{aa}  

\usepackage{graphicx}
\usepackage{txfonts}
\begin{document}

   \title{The SOPHIE search for northern extrasolar planets\thanks{Based on observations collected with the SOPHIE spectrograph on the 1.93m telescope at the Observatoire de Haute-Provence (CNRS), France, by the SOPHIE Consortium.}}

   \subtitle{XV. A Warm Neptune around the M-dwarf Gl378}

   \author{M. J. Hobson
          \inst{\ref{i:lam}}
          \and
          X. Delfosse\inst{\ref{i:grenoble}}
          \and
          N. Astudillo-Defru\inst{ \ref{i:chile}}
          \and
          I. Boisse\inst{\ref{i:lam}}
          \and
          R. F. D\'iaz\inst{\ref{i:uba},\ref{i:conicet}}
          \and
          F. Bouchy\inst{\ref{i:geneve}}
          \and
          X.Bonfils\inst{\ref{i:grenoble}}
          \and
          T. Forveille\inst{\ref{i:grenoble}}
          \and
          L. Arnold\inst{\ref{i:OHP}}
          \and
          S. Borgniet\inst{\ref{i:grenoble}}
          \and
          V. Bourrier\inst{\ref{i:geneve}}
          \and
          B. Brugger\inst{\ref{i:lam}}
          \and
          N. Cabrera Salazar\inst{\ref{i:grenoble}}
          \and
          B. Courcol\inst{\ref{i:lam}}
          \and
          S. Dalal\inst{\ref{i:paris}}
          \and
          M. Deleuil\inst{\ref{i:lam}}
          \and
          O. Demangeon\inst{\ref{i:porto}}
          \and
          X. Dumusque\inst{\ref{i:geneve}}
          \and
          N. Hara\inst{\ref{i:geneve}, \ref{i:ASD}}
          \and
          G. Hébrard\inst{\ref{i:paris}, \ref{i:OHP}}
          \and
          F. Kiefer\inst{\ref{i:paris}}
          \and
          T. Lopez\inst{\ref{i:lam}}
          \and
          L. Mignon\inst{\ref{i:grenoble}}
          \and
          G. Montagnier\inst{\ref{i:paris}, \ref{i:OHP}}
          \and
          O. Mousis\inst{\ref{i:lam}}
          \and
          C. Moutou\inst{\ref{i:lam},\ref{i:cfht}}
          \and
          F. Pepe\inst{\ref{i:geneve}}
          \and
          J. Rey\inst{\ref{i:geneve}}
          \and
          A. Santerne\inst{\ref{i:lam}}
          \and
          N. C. Santos\inst{\ref{i:porto},\ref{i:porto2}}
          \and
          M. Stalport\inst{\ref{i:geneve}}
          \and
          D. Ségransan\inst{\ref{i:geneve}}
          \and
          S. Udry\inst{\ref{i:geneve}}
          \and
          P.A Wilson\inst{\ref{i:warwick1},\ref{i:warwick2},\ref{i:paris}}
          }

   \institute{Aix Marseille Univ, CNRS, CNES, LAM, Marseille, France\\
              \email{melissa.hobson@lam.fr}\label{i:lam}
         \and
          Univ. Grenoble Alpes, CNRS, IPAG, 38000 Grenoble, France\label{i:grenoble}
         \and
         Universidad de Concepci\'on, Departamento de Astronom\'ia, Casilla 160-C, Concepci\'on, Chile\label{i:chile}
         \and
         Universidad de Buenos Aires, Facultad de Ciencias Exactas y Naturales. Buenos Aires, Argentina\label{i:uba}
         \and
         CONICET - Universidad de Buenos Aires. Instituto de Astronomía y Física del Espacio (IAFE). Buenos Aires, Argentina\label{i:conicet}        
         \and
         Observatoire Astronomique de l’Université de Genève, 51 Chemin des Maillettes, 1290 Versoix, Switzerland\label{i:geneve}
         \and
         Observatoire de Haute-Provence, CNRS, Aix Marseille Université, Institut Pythéas UMS 3470, 04870 Saint-Michel-l’Observatoire, France\label{i:OHP}
         \and
         Institut d’Astrophysique de Paris, UMR7095 CNRS, Université Pierre \& Marie Curie, 98bis boulevard Arago, 75014 Paris, France\label{i:paris}
         \and
         Instituto de Astrofísica e Ciências do Espaço, Universidade do Porto, CAUP, Rua das Estrelas, 4150-762 Porto, Portugal\label{i:porto}
         \and
         Departamento de Física e Astronomia, Faculdade de Ciências, Universidade do Porto, Rua do Campo Alegre, 4169-007 Porto, Portugal\label{i:porto2}
         \and
        ASD/IMCCE, CNRS-UMR8028, Observatoire de Paris,  PSL, UPMC, 77 Avenue Denfert-Rochereau, 75014 Paris, France \label{i:ASD}
         \and
         Canada-France-Hawaii Telescope Corporation, 65-1238 Mamalahoa Hwy, Kamuela, HI 96743, USA\label{i:cfht}
         \and
         Department of Physics, University of Warwick, Coventry CV4 7AL, UK\label{i:warwick1}
         \and
        Centre for Exoplanets and Habitability, University of Warwick, Coventry CV4 7AL, UK\label{i:warwick2}
        }

   \date{Received Dec 15, 2018; accepted Feb 12, 2019}

 
  \abstract
   {We present the detection of a Warm Neptune orbiting the M-dwarf Gl378, using radial velocity measurements obtained with the SOPHIE spectrograph at the Observatoire de Haute-Provence. The star was observed in the context of the SOPHIE exoplanets consortium's subprogramme dedicated to finding planets around M-dwarfs. Gl378 is an M1 star, of solar metallicity, at a distance of 14.96 pc. The single planet detected, Gl378 b, has a minimum mass of 13.02 $\rm M_{Earth}$ and an orbital period of 3.82 days, which place it at the lower boundary of the Hot Neptune desert. As one of only a few such planets around M-dwarfs, Gl378 b provides important clues to the evolutionary history of these close-in planets. In particular, the eccentricity of 0.1 may point to a high-eccentricity migration. The planet may also have lost part of its envelope due to irradiation.}


   \maketitle
%

\section{Introduction}

The mass-period diagram is an important diagnostic of the formation and evolution of planetary systems. There is a known dearth of Neptune-size exoplanets at short orbital periods compared to both Jupiter-size and Earth-size planets, which is generally referred to as the Neptune or sub-Jovian desert (\citealt{Lecavelier}, \citealt{Davis}, \citealt{Szabo}, \citealt{Beauge}, \citealt{Helled}, \citealt{Mazeh}). It is unlikely to be an observational bias, but is more probably due to photoevaporation and/or high-eccentricity migration (\citealt{Owen}, \citealt{Ionov}). 

Statistics on M-dwarf planets remain less certain than those on planets around Sun-type stars due to the comparatively small number of detections, though these are expected to increase thanks to several current or upcoming projects such as e.g. SPIRou \citep{Artigau}, CARMENES (e.g. \citealt{Quirrenbach14}, \citealt{Quirrenbach16}), HADES (e.g. \citealt{Affer}), and NIRPS \citep{Bouchy17} in radial velocity; or TESS (NASA mission, launched April 2018, \citealt{Ricker}), TRAPPIST (e.g. \citealt{Gillon}), SPECULOOS \citep{Delrez}, and ExTrA \citep{Bonfils15} in transits. Nevertheless, it is clear that while Hot Jupiters are rare around M-dwarfs, short-period Earths and superEarths are numerous, but Hot Neptunes remain unusual, making up only about 3\% of the sample of known exoplanets around M-dwarfs (e.g. \citealt{Bonfils}, \citealt{Dressing}, \citealt{Hirano}). 

The SOPHIE exoplanet consortium has led several ongoing exoplanet-hunting programmes on the SOPHIE spectrograph at the Observatoire de Haute Provence since 2006 (\citealt{Bouchy09}). Sub-programme 3, also known as SP3, is dedicated to the hunt for exoplanets around M-dwarf stars. Via a systematic survey of a volume-limited sample of M-dwarfs within 12 parsecs of the Sun, it seeks to detect superEarths and Neptunes, constrain the statistics of planets around M-dwarfs, and find potentially transiting companions. With a general radial velocity precision of 1-2 m/s on solar-type stars, SOPHIE has proved to be a successful planet hunter. For the SP3 in particular, we recently published the first two exoplanets from this sub-programme: the detection of Gl96 b and the independent confirmation of Gl617A b \citep{Hobson18}.

In this work, we report the detection of a Warm Neptune on the lower boundary of the Hot Neptune desert, orbiting the M-dwarf Gl378, which was observed as part of this survey. We describe the data and its analysis in Sect \ref{s: obs}  and Sect. \ref{s: dat an} respectively. The results are presented in Sect. \ref{s: res} and discussed in Sect. \ref{s: disc}. Finally, we conclude in Sect. \ref{s: Conc}.

\section{Observations} \label{s: obs}

Observations for Gl378 were gathered between 2015 and 2018 with the SOPHIE+ spectrograph (\citealt{Perruchot11}, \citealt{Bouchy13}). A total of 62 spectra were obtained. All the observations were performed with simultaneous sky measurement in order to check for potential moonlight contamination. Additionally, a ThAr or FP calibration spectrum was obtained immediately prior to each observation, in order to measure the instrumental drift (average value: 1 m/s). For the observations where the velocity difference between the moon and the star was less than 20 km/s, a merit function was applied, computed from the velocity difference and the S/N and CCF contrast in fibre B, in order to identify possible contamination. In this way, 18 observations were found to be contaminated by the moon and were discarded, leaving a total of 44 spectra. (We note that retaining these observations does not change the final planetary parameters within the uncertainties, but increases noise). With an exposure time of 1800s, the spectra have a median S/N of 83 (at 650 nm), resulting in a photon noise of $\rm \approx$ 3 m/s. For this star the photon noise is a little higher than that of instrument systematics (1-2 m/s).

\section{Data Analysis} \label{s: dat an}

The data were reduced using the SOPHIE Data Reduction Software (DRS, \citealt{Bouchy09}), which computes the radial velocity by cross-correlation functions. For M dwarfs this approach does not use all the Doppler content, so we extracted RVs through a template-matching algorithm. We shifted all the spectra to a common reference frame using the DRS RVs, and co-added them to build a high S/N stellar template. This template was Doppler shifted over a series of guess RVs, producing a Chi-square profile whose minimum corresponds to the maximum likelihood RV \citep{Astudillo15, Astudillo17c}.

\begin{figure}
    \centering
    \includegraphics[width=\hsize]{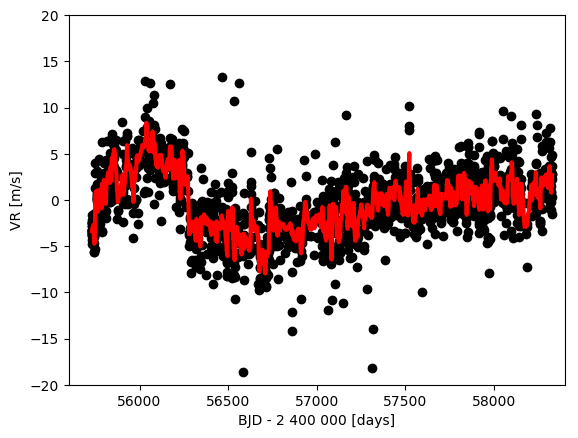}
    \caption{Correction for the long-term variations of the zero-point (red line) and the data points used to construct it (black dots). The points correspond to 11 stars, detailed in Sect. \ref{s: dat an}. The correction spans 7 years, with a dispersion of 2.87 m/s and a peak-to-peak variation of 16.3 m/s.}
    \label{fig:master}
\end{figure}

SOPHIE shows long-term variations of the zero-point, an effect first described in \cite{Courcol15}. We constructed an up-to-date correction from the SP3 stars plus the four solar-type 'super-constant' stars of the SOPHIE high-precision programmes as defined by \cite{Courcol15}, in the same way as in \cite{Hobson18}. Our updated constant correction uses 10 stars: the four super-constants, HD185144, HD221354, HD89269A, and HD9407; the three SP3 constants, Gl411, Gl514, Gl686; and the additional SP3 stars Gl521, Gl15A, and Gl694, selected because they have a corrected rms after the first iteration lower than 3 m/s (as defined by \citealt{Courcol15}). Fig. \ref{fig:master} shows the correction and the data used to derive it\footnote{This constant correction for SOPHIE RVs (applicable to M-dwarf stars) is available upon request.}.

The final radial velocities, which have a mean error bar of 3 m/s (including photon noise and instrumental error, following \citealt{Astudillo15}), are reported in Appendix \ref{A:RVs}, and were analysed with the Data and Analysis Center for Exoplanets (DACE) web platform\footnote{Available at \url{https://dace.unige.ch}}. We employed the keplerian-fitting tools (which follow \citealt{Delisle}) and the MCMC analysis facilities (described in \citealt{Diaz14} and \citealt{Diaz16}).

\subsection{Activity indicators}
\label{s:actind}

The main spectral activity indicators are the H alpha index, the log R'hk index, and the CCF bisector. We followed \cite{Boisse09} to compute the H alpha index; \cite{Astudillo17} to determine the log R'hk index from the Ca II H and K lines measured in the spectrum; and obtained the CCF bisector from the SOPHIE DRS. Since the Na I D lines have been shown to be good activity indicators for M-dwarfs (\citealt{Diaz07}, \citealt{Gomes11}), we also calculated the Na index as defined by \cite{Gomes11} from our SOPHIE spectra. The values obtained are given in Appendix \ref{A:RVs}.

\subsection{Stellar parameters}
\label{s:starparam}

The stellar parameters are listed in Table \ref{starprop}. Gl378 was characterized in \cite{Gaidos}, from where we obtained spectral type, magnitudes and colour indices (except for the K magnitude which was taken from \citealt{Cutri}), effective temperature, and luminosity. The coordinates, parallax, and distance were taken from the GAIA DR2 \citep{GAIAmission, GAIADR2}. We obtained the mean and standard deviation of $\rm log(R'_{HK})$ from the SOPHIE spectra, and used the $\rm log(R'_{HK})-log(P_{rot})$ relation from \cite{Astudillo17} to estimate a rotation period of $40.5 \pm 4$ days, with error bars calculated by propagation. We employed the MCAL code of \cite{Neves} to estimate the metallicity from our SOPHIE spectra. Finally, we used the distance measurement from \cite{GAIADR2} to estimate a more precise radius (following \citealt{Mann15}, with errors calculated by propagation) and stellar mass (using the MCMC routine provided by \citealt{Mann18}, based on masses from \citealt{Delfosse}) than were available in the literature.

\begin{table}[h]
\caption[]{\label{starprop}Stellar parameters.}
\begin{tabular}{lc}
\hline \hline
Parameter & Gl378 \\ \hline
Spectral Type   &       M1\tablefootmark{a}    \\
V       &       10.19\tablefootmark{a}  \\
B-V     &       1.26\tablefootmark{a}  \\
V-K     &       2.03\tablefootmark{b}  \\
Mass [$\rm M_\odot$] &      $0.56 \pm 0.01$\tablefootmark{c}      \\
Radius [$\rm R_\odot$] & $0.56 \pm 0.02$\tablefootmark{d}   \\
$\rm \alpha$ [h m s] & 10 02 21.7516441689\tablefootmark{e} \\ 
$\rm \delta$ [d m s] & +48 05 19.687165248\tablefootmark{e} \\
$\Pi$ [mas]     &       $66.8407 \pm 0.0322$\tablefootmark{e}      \\
Distance [pc]     &       $ 14.9609 \pm 0.0072$\tablefootmark{e}      \\
$\rm log(R'_{HK})$  &       $- 4.98 \pm 0.06$\tablefootmark{f}        \\
$\rm T_{eff}$ [k]   &       $3879 \pm 67$\tablefootmark{a}    \\
$\rm L_{\star}$ [$\rm L_\odot$] &       $0.06 \pm 0.01$\tablefootmark{a}             \\
Fe/H [dex]      &       $0.06 \pm 0.09$\tablefootmark{g}      \\
\hline
\end{tabular}
\tablefoot{ Sources:
\tablefoottext{a}{\cite{Gaidos}.}
\tablefoottext{b}{V: \cite{Gaidos}, K: \cite{Cutri}.}
\tablefoottext{c}{This work, following \cite{Mann18}.}
\tablefoottext{d}{This work, following \cite{Mann15}.}
\tablefoottext{e}{\cite{GAIADR2}.}
\tablefoottext{f}{This work, following \cite{Astudillo17}.}
\tablefoottext{g}{This work, following \cite{Neves}.}
}
\end{table}

\section{Results} \label{s: res}

\begin{figure}
    \centering
    \includegraphics[width=\hsize]{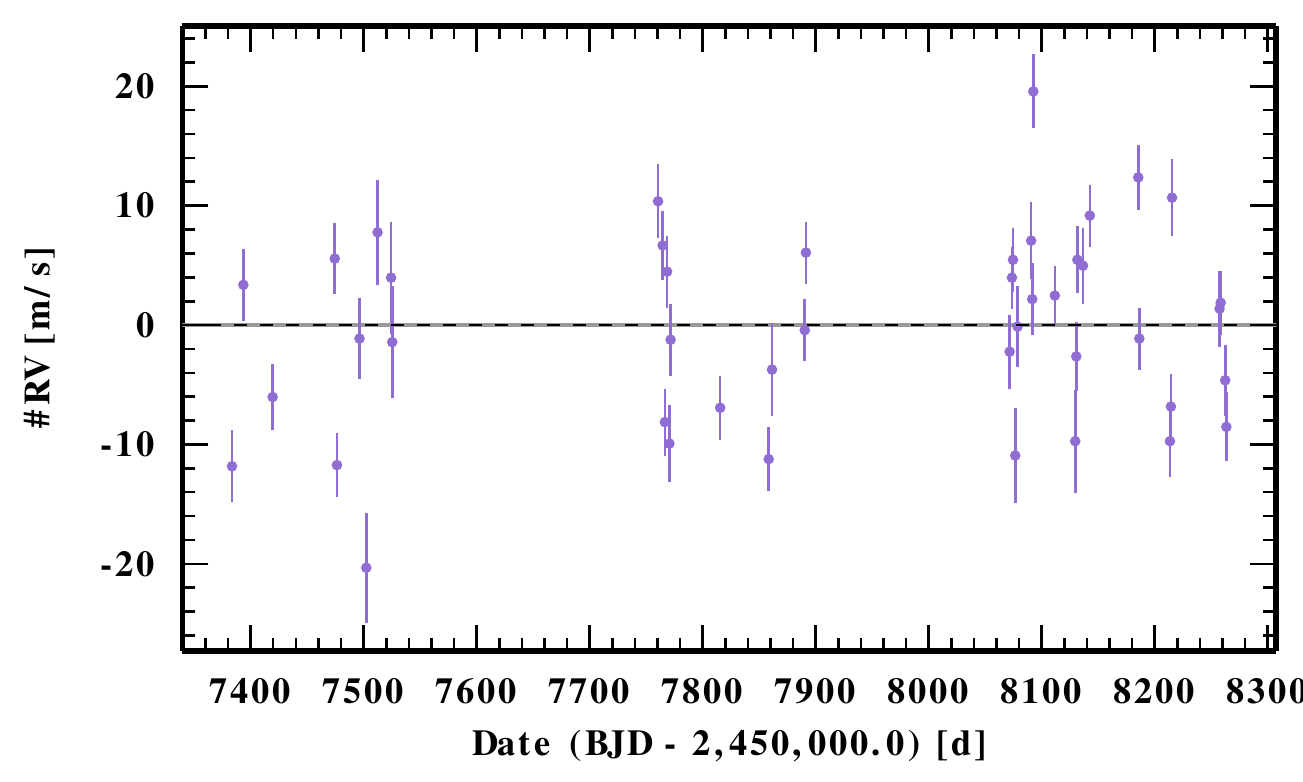}
    \caption{SOPHIE RVs for Gl378, obtained using template-matching, and corrected from the nightly drift and the long-term variations of the zero-point.}
    \label{fig:Gl378_dat}
\end{figure}

\begin{figure}
    \centering
    \includegraphics[width=\hsize]{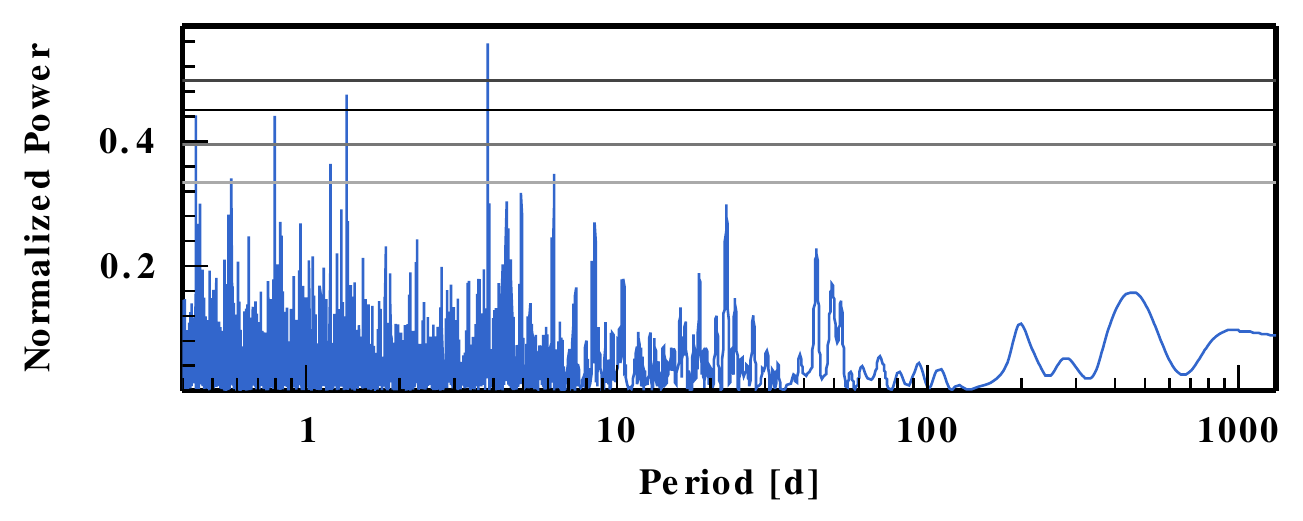}
    \caption{Periodogram of the SOPHIE RVs for Gl378, obtained using template-matching, and corrected from the nightly drift and the long-term variations of the zero-point. The horizontal lines indicate the 50\%, 10\%, 1\% and 0.1\% FAP levels respectively.}
    \label{fig:Gl378_per}
\end{figure}

\begin{figure}
    \centering
    \includegraphics[width=\hsize]{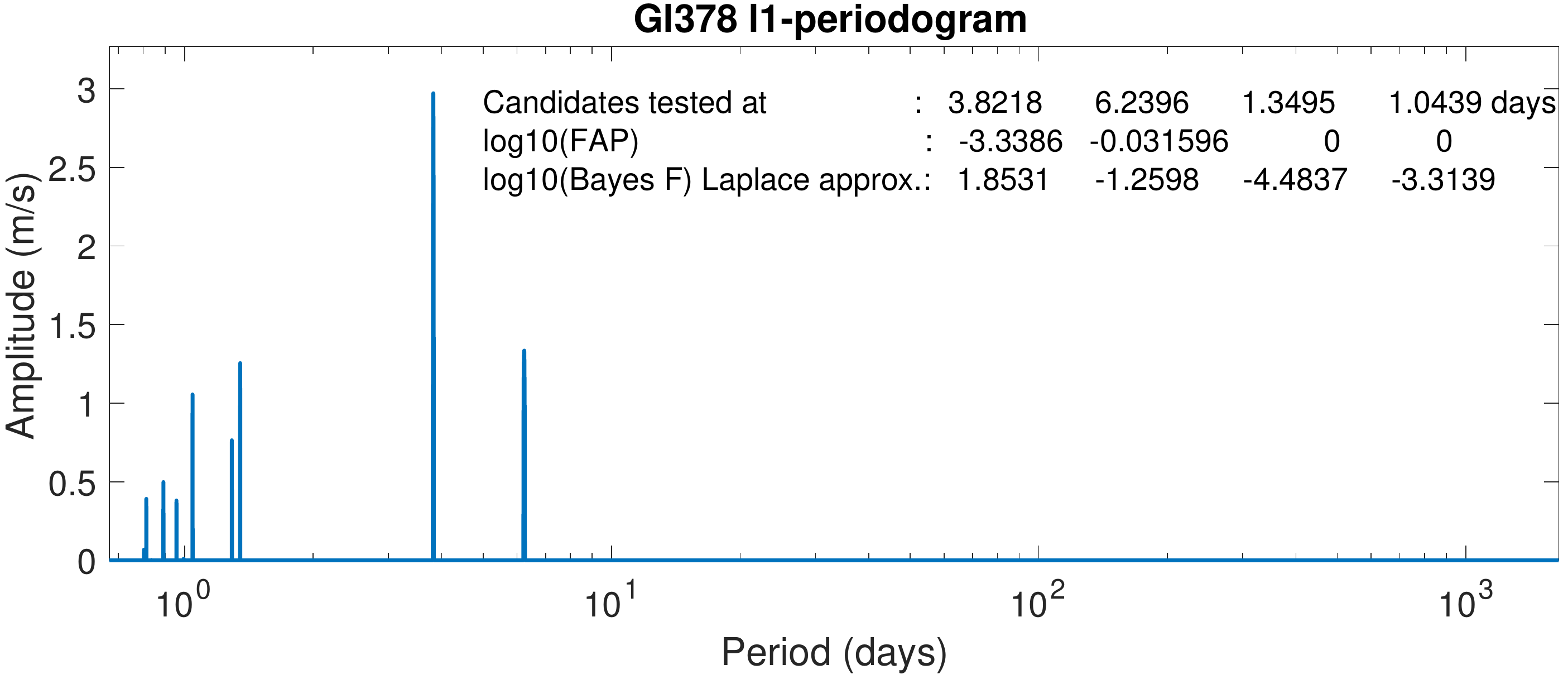}
    \caption{l1-Periodogram following \cite{Hara}. The FAP is computed according to \cite{Baluev08}'s formula, and the Bayes Factor is computed via a Laplace approximation with the same methodology as in \cite{Nelson}, appendix A.4.}
    \label{fig:Gl378_per_l1}
\end{figure}

The time series of the radial velocities is shown in Fig.  \ref{fig:Gl378_dat}, and its periodogram in Fig. \ref{fig:Gl378_per}. The periodogram shows a clear peak at 3.82d, which by bootstrap resampling we place below 0.01\% false alarm probability (FAP). The other notable peaks, at 0.79d and 1.35d, are 1-day aliases of the 3.82d period; they are systematically weaker than the 3.82d peak, and attempted Keplerian fits have higher $\sigma_{(O-C)}$. We also computed the l1-periodogram as in \cite{Hara}, which is shown in Fig. \ref{fig:Gl378_per_l1}, and confirms the 3.82d signal as the most significant. 

\begin{figure}
    \centering
    \includegraphics[width=\hsize]{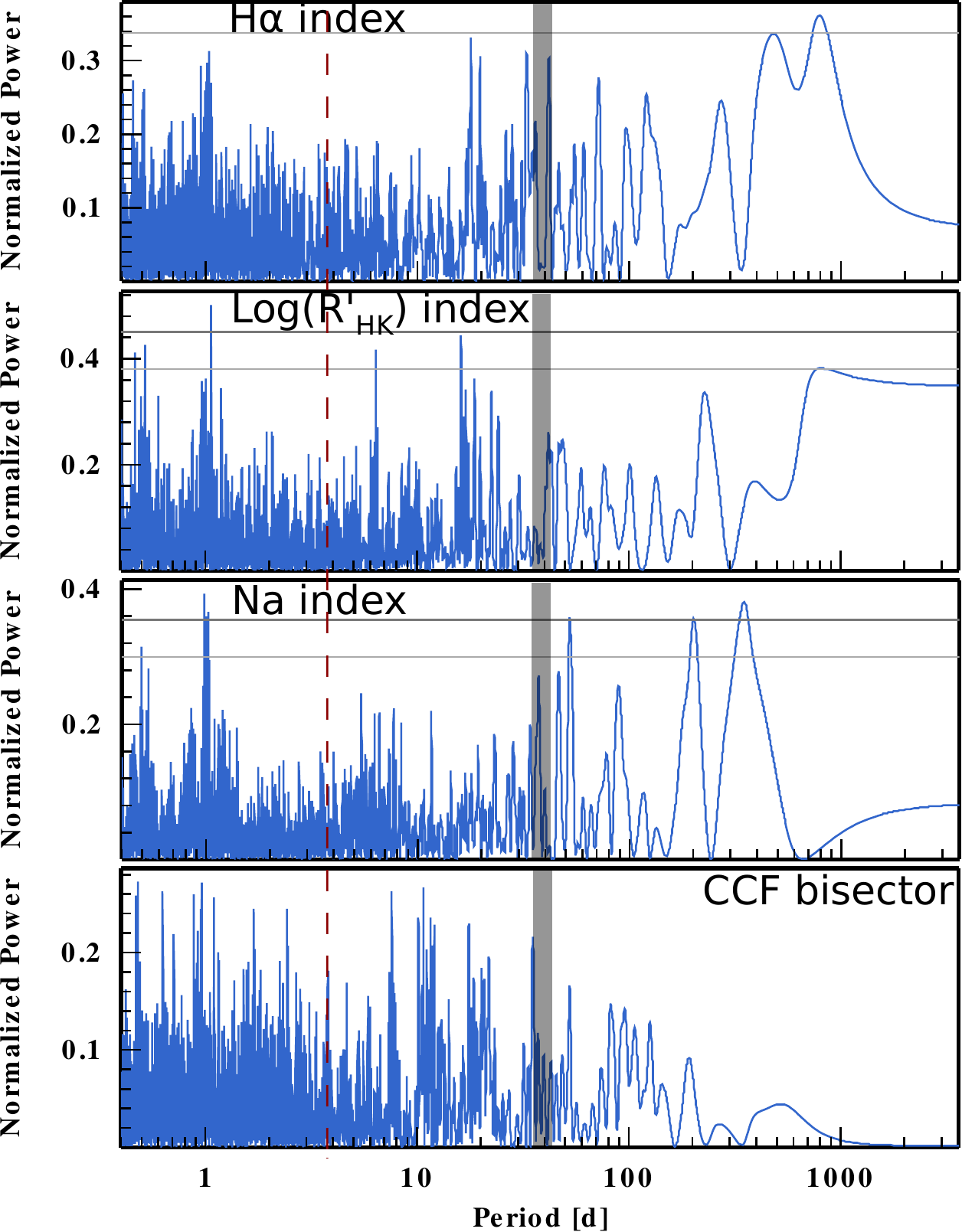}
    \caption{Periodogram of the activity indices for Gl378: from top to bottom, H$\alpha$ index, log($R'_{HK}$) index, Na index, and CCF bisector. The horizontal lines indicate the 50\% and 10\% FAP levels respectively. The red vertical dashed line marks the orbital period of Gl378 b. The shaded grey region indicates the probable rotation period, as estimated from the log($R'_{HK}$).}
    \label{fig:Gl378_act}
\end{figure}

The periodograms of the four activity indicators described in Sect. \ref{s:actind} are shown in Fig. \ref{fig:Gl378_act}. None of them show any peaks below 10\% FAP, or any peak whatsoever at the 3.82d period found in the RVs. Likewise, there is no anticorrelation in evidence between the RVs and the CCF bisector. Additionally, for an M-dwarf, a rotation period of 3.82d would lead to an extremely high activity level, with saturated chromospheric emission of H$\rm \alpha$ \citep{Delfosse98} and Ca \citep{Astudillo17}, which is clearly not the case for Gl378. Finally, \cite{Houdebine} found Gl378 to be a slow rotator, with $v \sin{i} = 2.25$ km/s. Therefore, we conclude that the 3.82d peak cannot be of stellar origin, and that Gl378 shows no evidence of clear stellar activity (at the estimated rotation period of $\rm P_{Rot} = 40.5 \pm 4 $ d, indicated by the shaded grey regions in Fig. \ref{fig:Gl378_act}, or any other period) in the SOPHIE spectra.

\begin{figure}
    \centering
    \includegraphics[width=\hsize]{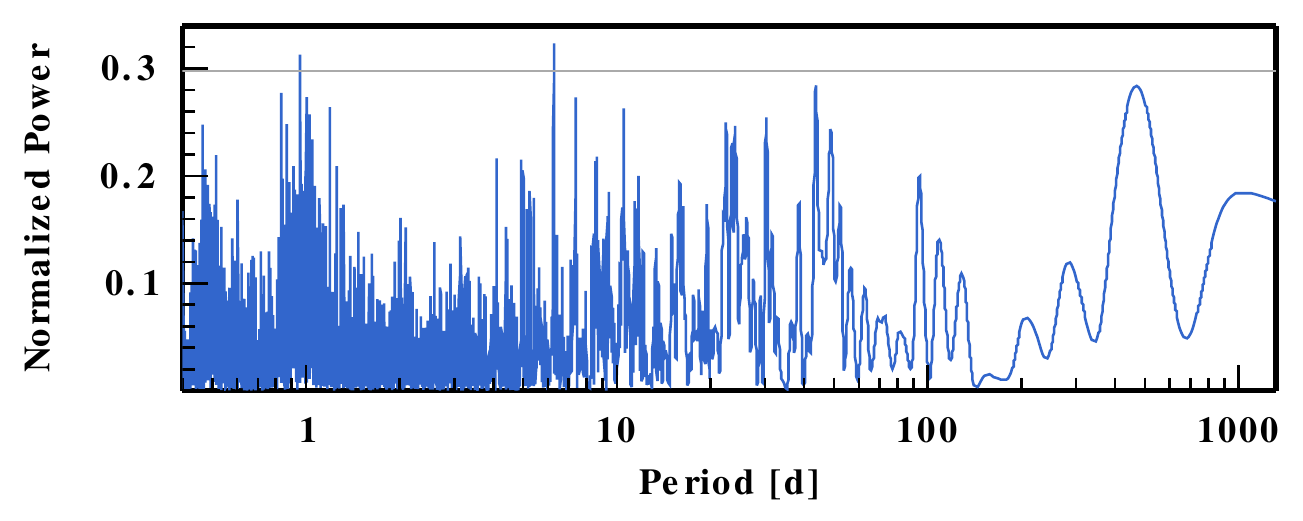}
    \caption{Periodogram of the residuals of the SOPHIE RVs for Gl378 following the fit of a planet at 3.82 d. The horizontal line indicates the 50\% FAP level.}
    \label{fig:Gl378_per_res}
\end{figure}

\begin{figure}
    \centering
    \includegraphics[width=\hsize]{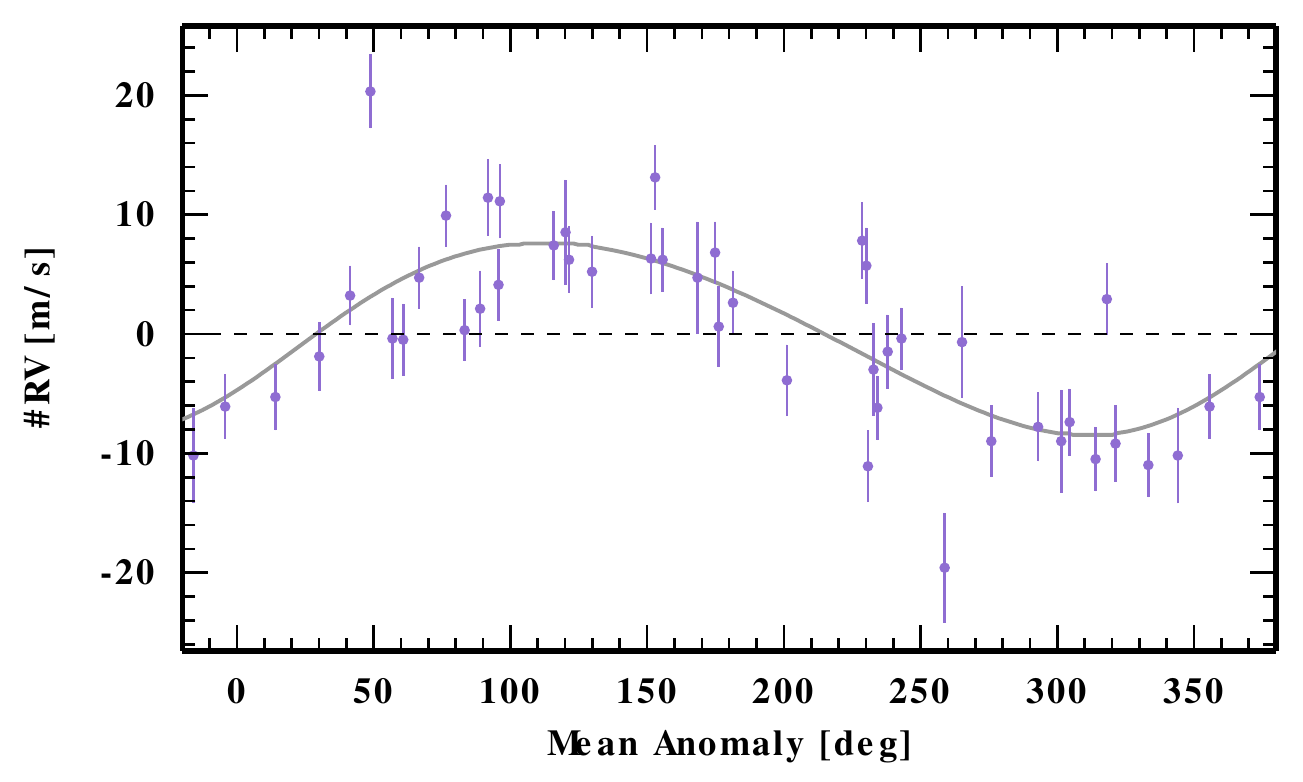}
    \caption{SOPHIE RVs for Gl378, phase-folded to the one-planet model with P=3.82d. The curve indicates the fitted model.}
    \label{fig:Gl378_dat_phasefold}
\end{figure}

We employed the DACE platform to fit a Keplerian signal to the 3.82d period. The highest peak in the periodogram of the residuals has a FAP of 17\%, and is therefore not significant (Fig. \ref{fig:Gl378_per_res}). In order to sample the joint posterior distribution of the model parameters, we proceeded to carry out an MCMC analysis. We used a model with a single keplerian and an additive stellar jitter. The resulting parameters are summarised in Table \ref{Gl378_tab-mcmc-Summary_params}, with the full outputs available in Appendix \ref{A:MCMC}. The phase-folded data and fitted keplerian are shown in Fig. \ref{fig:Gl378_dat_phasefold}.

\begin{table}
\caption{Best-fit solution for the planetary system orbiting Gl378.}  \label{Gl378_tab-mcmc-Summary_params}
\begin{tabular}{lcc}
\hline
\hline
Param. & Units & Gl378 b \\
\hline \\[-1.5ex]
$P$ & [d] & 3.822$_{-0.001}^{+0.001}$  \\ [0.5ex]
$K$ & [m\,s$^{-1}$] & 7.96$_{-1.23}^{+1.24}$  \\[0.5ex]
$e$ &   & 0.109$_{-0.077}^{+0.131}$  \\[0.5ex]
$\omega$ & [deg] & 210.6$_{-116.9}^{+79.8}$  \\[0.5ex]
$T_P$ & [d] & 2455500.02$_{-1.3}^{+1.27}$  \\[0.5ex]
$T_C$ & [d] & 2455502.564$_{-1.05}^{+0.72}$  \\[0.5ex]
\hline \\[-1.5ex]
$a$ & [AU] & 0.039435$_{-0.00023}^{+0.00023}$  \\[0.5ex]
M.$\sin{i}$ & [M$_{\rm Earth}$] & 13.02$_{-2.01}^{+2.03}$  \\[0.5ex]
\hline \\[-1.5ex]
$\gamma_{SOPHIE}$ & [m\,s$^{-1}$] & \multicolumn{1}{c}{-9697.333$_{-0.83}^{+0.85}$}\\[0.5ex]
$\sigma_{JIT}$ & [m\,s$^{-1}$] & \multicolumn{1}{c}{4.610$_{-0.74}^{+0.83}$}\\[0.5ex]
$\sigma_{(O-C)}$ & [m\,s$^{-1}$] & \multicolumn{1}{c}{4.86}\\[0.5ex]
$\log{(\rm Likelihood})$ &   & \multicolumn{1}{c}{-138.07$_{-2.09}^{+1.41}$}\\[0.5ex]
\hline
\end{tabular}
\tablefoot{ For each parameter the median of the posterior is reported, with error bars computed from the MCMC chains using a 68.3\% confidence interval. $\sigma_{O-C}$ corresponds to the weighted standard deviation of the residuals around the best solutions. All the parameters probed by the MCMC can be found in Appendix \ref{A:MCMC}, Table \ref{tab:GL378_mcmc-allparams}.}
\end{table}

\section{Discussion} \label{s: disc}

\begin{figure}
    \centering
    \includegraphics[width=\hsize]{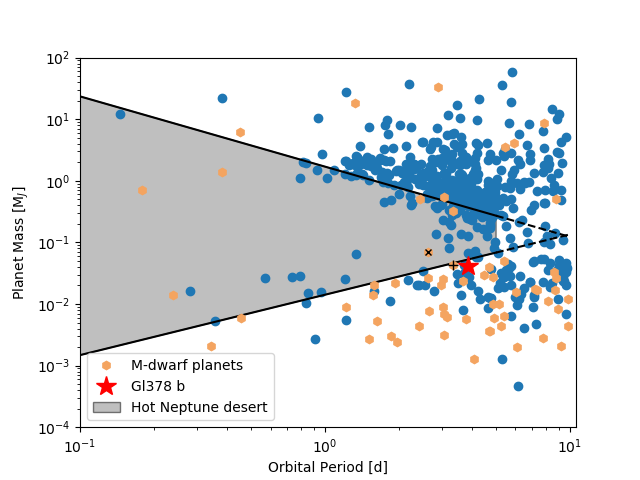}
    \caption{The Hot Neptune desert. Gl378 b is indicated by the red star. All planets with known masses and periods<10d are shown (taken from exoplanets.eu on 18 Oct 2018). Planets orbiting M-dwarfs are highlighted (orange hexagons), with GJ436 b and GJ3470 marked by $\times$ and $+$ symbols respectively. The location of the Hot-Neptune desert is indicated (grey region), using the boundaries of \cite{Mazeh} (black lines). The boundaries are dotted beyond 5d to reflect \cite{Mazeh}'s warning that beyond this orbital period the existence of the desert is uncertain.}
    \label{fig:HNdes}
\end{figure}

With a minimum mass of 13.63 M$_{\rm Earth}$ and an orbital period of 3.82 d, Gl378 b is a Warm Neptune-like exoplanet. Depending on the heat redistribution factor and albedo assumed, we obtain equilibrium temperatures in the range of $\rm T_{eq} \approx 630 K$ (for a heat redistribution factor of 1 and an albedo of 0.3, close to that of Neptune) to $\rm T_{eq} \approx 830 K$ (for a redistribution factor of 2 and an albedo of 0, as a lower limit). Its orbital parameters place it on the lower boundary of the Neptune desert as defined by \cite{Mazeh}, as shown in Fig. \ref{fig:HNdes} (although they note the lower boundary is somewhat blurry, and other authors such as e.g. \citealt{Owen} have placed the external limit in period at $\rm P \approx 3d$ rather than $\rm P \approx 5d$). This location on the lower boundary rather than within the desert, and its range of probable equilibrium temperature well below the methane condensation temperature of 1200K, lead us to class it as a Warm rather than a Hot Neptune. The lower boundary is believed to have its origin in photoevaporation (\citealt{Owen}). Therefore, Gl378 b may have lost at least part of its gaseous envelope due to the X-Ray and EUV irradiation from its host star. We note that the young active phase is long for M dwarfs compared to Sun-type stars, giving more time for evaporation to work.

Gl378 b is likely similar - assuming its mass is close to its M.$\sin{i}$ - to GJ436 b (\citealt{Ehrenreich}, \citealt{Lavie}) and GJ3470 b \citep{Bourrier18b}. These planets are Warm Neptunes, in the same region of Fig. \ref{fig:HNdes} as Gl378 b, and orbit M-dwarf stars. Both are surrounded by giant hydrogen exospheres; GJ436b possibly became a Warm Neptune recently due to a late high-eccentricity migration and is thus not losing much mass \citep{Bourrier15, Bourrier16, Bourrier18a}, while GJ3470b is much more irradiated by its younger and earlier-type star and could have lost up to 35\% of its mass already \citep{Bourrier18b}. This suggests that the Warm Neptune population at the border of the desert is particularly sensitive to atmospheric escape, and supports this mechanism as the reason why Hot Neptunes are missing. The three planets also all have similar non-zero eccentricities (\citealt{Deming}, \citealt{Kosiarek}), which may point to high-eccentricity rather than disk-driven migrations. Therefore, objects like Gl378 b are crucial to understanding the evolution of close-in planets, providing we can characterize them.

A fuller characterization of the planet would require knowledge of its density, and therefore of its radius. The transit probability of a planet detected by radial velocity can be approximated by $\rm P(transit) \approx R_{\star}/a$, with $R_{\star}$ the stellar radius and $a$ the semi-major axis of the planetary orbit (\citealt{Borucki}). For Gl378b, we obtain a transit probability of $\rm P(transit) = 6.5 \pm 0.5 \%$. As Gl378 b is most likely a Neptune-mass exoplanet, the analysis of \cite{Stevens} suggests this transit probability is likely underestimated: they derive the posterior transit probability from the prior distributions of planetary masses and inclinations, finding that physically motivated distributions from planet formation models yield increased posterior transit probabilities for Neptune-mass planets. The transit depth is given by $\rm \Delta F = {(R_p / R_{\star})}^2$. Using the mass-radius relation of \cite{Chen}, we estimate the radius of Gl378 b as $\rm R_p \approx 4.67 R_{\oplus}$, which when combined with the stellar radius gives a transit depth of $\rm \Delta F \approx 0.58 \%$. This could be observed by ground-based surveys, and should be easily detectable by space-based missions such as TESS (\citealt{Barclay}) or CHEOPS (\citealt{CHEOPS}).

The metallicity estimated for the host star, Gl378, from our SOPHIE spectra is of $\rm [Fe/H] = 0.06 \pm 0.09$. This solar metallicity is in line with the tendency for M-dwarfs hosting planets to be comparatively more metal-rich (e.g. \citealt{Courcol16}, \citealt{Hirano}, who find $\rm [Fe/H]\gtrsim 0$ for planet hosts) with respect to the sub-solar average metallicity of nearby M-dwarfs (e.g. \citealt{Schlaufman} who report $\rm [Fe/H] = -0.17 \pm 0.07$, \citealt{Passegger} who find mainly subsolar values for the CARMENES target sample); and also with the fact that Hot Neptunes are more common with increasing metallicity (\citealt{Petigura} - although we note this is a study on the Kepler survey, and therefore the planets are categorized by radius rather than mass).

The activity indicators analyzed show no evidence for quasi-periodic stellar activity signals. Likewise, the residuals of the keplerian fit to the 3.82 d planet show no significant periodicity, although statistics on M-dwarfs indicate that most of their planets are found in multi-planet systems (e.g. \citealt{Bonfils}, \citealt{Dressing}). However, the $\sigma_{O-C}$ (weighted standard deviation of the residuals around the best solution) is high compared to the mean error bar of the observations, suggesting there are further effects in the data - additional planets, stellar activity signals, and/or systematics. As an M-dwarf, Gl378 is faint in the visible, emitting most of its radiation in the infrared; consequently, high-precision infrared spectroscopy could help to detect further planets, or place limits on their existence.

\section{Conclusions} \label{s: Conc}

We have presented the detection of a Neptune-mass planet orbiting the M-dwarf Gl378 at a period of 3.82d. Its orbital parameters place it at the edge of the Hot Neptune desert, though with a $sin(i)$ degeneracy on the mass. Transit observations could help to break this degeneracy: if the planet transits, the inclination can be measured, while a non-detection can place limits on it. Transit measurements would also provide the radius and therefore the density, permitting a characterization of its probable composition. Finally, if the planet transits, we should able to characterize its atmosphere in the UV/Ly-$\rm \alpha$ line, given the brightness and close distance of its host star to us, and the fact the planet is a warm Neptune and therefore likely surrounded by a giant hydrogen exosphere. This exosphere can extend beyond the Roche lobe, resulting in an effective planet radius in UV an order of magnitude larger than the optical radius, and hence a slightly increased transit probability. Therefore, UV/Ly-$\rm \alpha$ observations at the expected transit times would be interesting even if the planet does not transit in the optical.

Although we do not detect any other periodic signals in our data, it is statistically likely that more planets are present. Monitoring this star with infrared spectroscopy should help to resolve this question, and to refine the ephemeris of Gl378b for transit searches. We hope to observe Gl378 with SPIRou in the near future.

\begin{acknowledgements}
We warmly thank the OHP staff for their support on the observations.
We thank the anonymous referee for their careful reading and valuable comments which helped to improve the manuscript.
This work was supported by the Programme National de Plan\'etologie (PNP) of CNRS/INSU, co-funded by CNE.
X.De., X.B., I.B. and T.F. received funding from the French Programme National de Physique Stellaire (PNPS) and the Programme National de Plan\'etologie (PNP) of CNRS (INSU).
X.B. acknowledges funding from the European Research Council under the ERC Grant Agreement n. 337591-ExTrA. 
This work has been supported by a grant from Labex OSUG@2020 (Investissements d'avenir – ANR10 LABX56).
This work is also supported by the French National Research Agency in the framework of the Investissements d’Avenir program (ANR-15-IDEX-02), through the funding of the "Origin of Life" project of the Univ. Grenoble-Alpes.

V. B. acknowledges support by the Swiss National Science Foundation (SNSF) in the frame of the National Centre for Competence in Research PlanetS, and has received funding from the European Research Council (ERC) under the European Unions Horizon 2020 research and innovation programme (project Four Aces; grant agreement No 724427). 

This work was supported by FCT - Fundação para a Ciência e a Tecnologia through national funds and by FEDER through COMPETE2020 - Programa Operacional Competitividade e Internacionalização by these grants: UID/FIS/04434/2013 \& POCI-01-0145-FEDER-007672; PTDC/FIS-AST/28953/2017 \& POCI-01-0145-FEDER-028953 and PTDC/FIS-AST/32113/2017 \& POCI-01-0145-FEDER-032113.

N.A-D. acknowledges support from FONDECYT \#3180063.

N. H. acknowledges the financial support of the National Centre for Competence in Research PlanetS of the Swiss National Science Foundation (SNSF).

X.Du. is grateful to the Branco Weiss Fellowship—Society in Science for continuous support.

This publication makes use of The Data \& Analysis Center for Exoplanets (DACE), which is a facility based at the University of Geneva (CH) dedicated to extrasolar planets data visualisation, exchange, and analysis. DACE is a platform of the Swiss National Centre of Competence in Research (NCCR) PlanetS, federating the Swiss expertise in Exoplanet research. The DACE platform is available at https://dace.unige.ch.

\end{acknowledgements}

\bibliographystyle{aa} 
\bibliography{biblio} 

\begin{appendix}
\section{Radial velocities}
\label{A:RVs}

In this appendix, we present the radial velocities obtained for Gl378 from SOPHIE+ using template-matching, corrected from the nightly drift and the long-term variations of the zero-point.

\onecolumn

\begin{longtable}{llllllllll}
\caption{\label{Gl378_RV_t} Radial velocities for Gl378.}\\ 
\hline\hline
BJD $[-2400000 d]$ & RV $[km/s]$ & $\rm \sigma_{RV}$ $[km/s]$ & Bisector $[m/s]$ & $\rm H \alpha$ index & $\rm \sigma_{H \alpha}$  & $\rm log(R'_{HK})$  & $\rm \sigma_{log(R'_{HK})}$ & Na index & $\rm \sigma_{Na}$  \\
\hline
\endfirsthead
\caption{continued.}\\
\hline\hline
BJD $[-2400000 d]$ & RV $[km/s]$ & $\rm \sigma_{RV}$ $[km/s]$ & Bisector $[m/s]$ & $\rm H \alpha$ index & $\rm \sigma_{H \alpha}$  & $\rm log(R'_{HK})$  & $\rm \sigma_{log(R'_{HK})}$ & Na index & $\rm \sigma_{Na}$  \\
\hline
\endhead
\hline
\endfoot
57383.6037 & -9.7084 & 0.003  & 15.167 & 0.2353 & 0.0012 & -4.9599 & 0.0001 & 0.2107 & 0.0103 \\
57393.6361 & -9.6932 & 0.003  & 11.333 & 0.2242 & 0.0012 & -4.9868 & 0.0001 & 0.2116 & 0.0101 \\
57419.5259 & -9.7026 & 0.0028 & -2     & 0.2255 & 0.0011 & -4.9107 & 0.0001 & 0.2085 & 0.0094 \\
57474.4968 & -9.691  & 0.003  & 20.667 & 0.2236 & 0.0012 & -5.0253 & 0.0001 & 0.2079 & 0.0100 \\
57476.428  & -9.7083 & 0.0027 & 21.333 & 0.2248 & 0.0010 & -5.002  & 0.0001 & 0.2045 & 0.0089 \\
57496.4268 & -9.6977 & 0.0034 & 16.333 & 0.2343 & 0.0015 & -4.9589 & 0.0001 & 0.2163 & 0.0110 \\
57502.3933 & -9.7169 & 0.0046 & -8.667 & 0.2386 & 0.0021 & -4.9313 & 0.0001 & 0.2212 & 0.0145 \\
57512.3882 & -9.6888 & 0.0044 & 13     & 0.2267 & 0.0019 & -4.9842 & 0.0001 & 0.2115 & 0.0139 \\
57524.3675 & -9.6926 & 0.0047 & -1     & 0.2389 & 0.0022 & -5.0567 & 0.0001 & 0.2199 & 0.0156 \\
57525.3946 & -9.698  & 0.0047 & -1.333 & 0.2416 & 0.0022 & -5.0498 & 0.0001 & 0.214  & 0.0150 \\
57760.5835 & -9.6862 & 0.0031 & 15.167 & 0.2227 & 0.0013 & -4.9891 & 0.0001 & 0.217  & 0.0104 \\
57764.615  & -9.6899 & 0.0029 & 7.167  & 0.2223 & 0.0012 & -4.9965 & 0.0001 & 0.2174 & 0.0101 \\
57766.6183 & -9.7047 & 0.0028 & 5.333  & 0.2258 & 0.0011 & -5.0322 & 0.0001 & 0.2127 & 0.0010 \\
57768.5861 & -9.6921 & 0.003  & 8.167  & 0.2228 & 0.0012 & -4.9964 & 0.0001 & 0.2254 & 0.0105 \\
57770.6187 & -9.7065 & 0.0032 & 3      & 0.2263 & 0.0013 & -5.0235 & 0.0001 & 0.2221 & 0.0111 \\
57771.6751 & -9.6978 & 0.003  & 18     & 0.2226 & 0.0012 & -5.0034 & 0.0001 & 0.2163 & 0.0101 \\
57815.5618 & -9.7035 & 0.0027 & 11.833 & 0.222  & 0.0011 & -5.0024 & 0.0001 & 0.2122 & 0.0093 \\
57858.4545 & -9.7078 & 0.0027 & 10.333 & 0.232  & 0.0010 & -4.996  & 0.0001 & 0.2051 & 0.0092 \\
57861.4145 & -9.7003 & 0.0039 & 19.5   & 0.2341 & 0.0017 & -5.0434 & 0.0001 & 0.2229 & 0.0129 \\
57890.4053 & -9.697  & 0.0026 & 13.5   & 0.2314 & 0.0009 & -4.9607 & 0.0001 & 0.1997 & 0.0084 \\
57891.3779 & -9.6905 & 0.0026 & 10     & 0.2285 & 0.0009 & -4.9447 & 0.0001 & 0.2038 & 0.0086 \\
58071.6969 & -9.6988 & 0.0031 & 15     & 0.2358 & 0.0012 & -4.9848 & 0.0001 & 0.2123 & 0.0104 \\
58073.6999 & -9.6926 & 0.0026 & 9.833  & 0.2326 & 0.0010 & -4.9674 & 0.0001 & 0.2074 & 0.0092 \\
58074.6453 & -9.6911 & 0.0027 & 21     & 0.2421 & 0.0010 & -4.9242 & 0.0001 & 0.2085 & 0.0094 \\
58076.6457 & -9.7075 & 0.004  & 3.167  & 0.233  & 0.0018 & -4.9704 & 0.0001 & 0.2162 & 0.0130 \\
58078.6855 & -9.6967 & 0.0034 & 2      & 0.2365 & 0.0014 & -4.9643 & 0.0001 & 0.2133 & 0.0110 \\
58090.7094 & -9.6895 & 0.0032 & 3.333  & 0.2317 & 0.0014 & -4.941  & 0.0001 & 0.2131 & 0.0108 \\
58091.6594 & -9.6944 & 0.003  & 4.333  & 0.2339 & 0.0012 & -4.9547 & 0.0001 & 0.2121 & 0.0102 \\
58092.6222 & -9.677  & 0.0031 & 7.833  & 0.2373 & 0.0013 & -4.9128 & 0.0001 & 0.2173 & 0.0107 \\
58111.6548 & -9.6941 & 0.0025 & 15     & 0.254  & 0.0010 & -4.8508 & 0.0001 & 0.2127 & 0.0091 \\
58129.7054 & -9.7063 & 0.0043 & 0.333  & 0.2243 & 0.0019 & -4.9616 & 0.0001 & 0.2457 & 0.0146 \\
58130.6467 & -9.6992 & 0.0029 & 14.667 & 0.2311 & 0.0011 & -4.9647 & 0.0001 & 0.2151 & 0.0010 \\
58131.6165 & -9.6911 & 0.0028 & 3.333  & 0.2299 & 0.0011 & -4.9829 & 0.0001 & 0.211  & 0.0097 \\
58136.5933 & -9.6916 & 0.0032 & 8.833  & 0.2296 & 0.0014 & -4.9599 & 0.0001 & 0.2241 & 0.0111 \\
58142.6058 & -9.6874 & 0.0026 & 11.167 & 0.2363 & 0.0010 & -4.9272 & 0.0001 & 0.2125 & 0.0094 \\
58185.4633 & -9.6842 & 0.0027 & 4.833  & 0.2277 & 0.0010 & 999.99  & 0.0001 & 0.215  & 0.0093 \\
58186.4198 & -9.6977 & 0.0026 & 8      & 0.2313 & 0.0010 & 999.99  & 0.0001 & 0.2103 & 0.0089 \\
58213.5248 & -9.7063 & 0.003  & 15.833 & 0.2405 & 0.0011 & 999.99  & 0.0001 & 0.2094 & 0.0097 \\
58214.3723 & -9.7034 & 0.0027 & 11.833 & 0.2334 & 0.0010 & 999.99  & 0.0001 & 0.2095 & 0.0092 \\
58215.3927 & -9.6859 & 0.0032 & 14.833 & 0.2378 & 0.0013 & 999.99  & 0.0001 & 0.2147 & 0.0103 \\
58257.4077 & -9.6952 & 0.0032 & 10.667 & 0.2226 & 0.0012 & 999.99  & 0.0001 & 0.2029 & 0.0099 \\
58258.3895 & -9.6947 & 0.0027 & 5.5    & 0.2254 & 0.0010 & 999.99  & 0.0001 & 0.2029 & 0.0090 \\
58262.4213 & -9.7012 & 0.003  & 11.5   & 0.2231 & 0.0011 & 999.99  & 0.0001 & 0.2035 & 0.0097 \\
58263.3966 & -9.7051 & 0.0029 & 7.667  & 0.2248 & 0.0011 & 999.99  & 0.0001 & 0.2006 & 0.0093
\end{longtable}

\clearpage

\section{MCMC parameters}
\label{A:MCMC}

In this appendix, we present the parameters probed by the MCMC analysis that was applied to the radial velocities of Gl378.

\begin{sidewaystable*}
\tiny
  \begin{center}
    \caption{Parameters probed by the MCMC used to fit the RV measurements of GL378.
      The maximum likelihood solution, median, mode,
      and standard-deviation of the posterior distribution
      for each parameter are shown, as well as the 68.27\%,
      95.45\%, and 99.73\% confidence intervals. Parameters with priors listed are fitting parameters, while the rest are derived.
      The prior for each parameter can be of type: $\mathcal{U}$:~uniform,
      $\mathcal{N}$:~normal, or $\mathcal{TN}$:~truncated normal. Priors without given ranges are improper. The mean longitude $\lambda_0$ is given at reference epoch: 2455500.0~BJD.}
    \label{tab:GL378_mcmc-allparams}
    \resizebox{!}{0.95\height}{
    \begin{tabular}{cccccccccccc}
      \hline
      Parameter & Units & Max(Likelihood) & Mode & Mean & Std & Median & 68.27\% & 95.45\% & 99.73\% & Prior\\
      \hline
      $\log$(Likelihood) &  & -135.12 & -137.44 & -138.41 & 1.85 & -138.07 & [-140.16--136.66] & [-143.08--135.84] & [-147.57--135.37] & --\\
        \hline
        \multicolumn{11}{c}{\textbf{Star}}\\
        \hline
      $M_\mathrm{{S}}$ & [$M_\odot$] & 0.5458 & 0.5591 & 0.5600 & 0.0100 & 0.5599 & [0.5500-0.5699] & [0.5399-0.5801] & [0.5297-0.5900] & $\mathcal{{U}}$\\
      $\Pi_\mathrm{{S}}$ & [mas] & 66.8313 & 66.8364 & 66.8405 & 0.0322 & 66.8402 & [66.8083-66.8721] & [66.7762-66.9052] & [66.7465-66.9380] & $\mathcal{{U}}$\\
      \hline
      \multicolumn{11}{c}{\textbf{Offset}}\\
      \hline
      $\gamma_{\mathrm{NAIRA(DRS-errinc3)}}$ & [m/s] & -9697.488 & -9697.317 & -9697.329 & 0.864 & -9697.333 & [-9698.165--9696.485] & [-9699.075--9695.588] & [-9700.136--9694.614] & $\mathcal{{U}}$\\
      \hline
      \multicolumn{11}{c}{\textbf{Noise}}\\
      \hline
      Avg. act. & [m/s] & 4.073 & 4.521 & 4.664 & 0.806 & 4.610 & [3.874-5.441] & [3.199-6.417] & [2.658-7.737] & $\mathcal{U}(0,20)$\\
          \hline
          \multicolumn{11}{c}{\textbf{GL378b}}\\
          \hline
      $P$ & [d] & 3.82264 & 3.82248 & 3.82249 & 0.00135 & 3.82248 & [3.82119-3.82381] & [3.81973-3.82522] & [3.81806-3.82698] & $\mathcal{{U}}$\\
      $K$ & [m/s] & 8.16 & 7.91 & 7.96 & 1.24 & 7.96 & [6.73-9.20] & [5.51-10.47] & [4.19-11.95] & $\mathcal{{U}}$\\
      $e$ &  & 0.098 & 0.000 & 0.132 & 0.103 & 0.109 & [0.032-0.234] & [0.004-0.393] & [0.000-0.595] & $\mathcal{{U}}$\\
      $\omega$ & [$^\circ$] & 218.4 & 244.6 & 197.0 & 90.7 & 210.6 & [93.7-290.4] & [13.1-346.8] & [0.8-359.4] & $\mathcal{{U}}$\\
      $\lambda_0$ & [$^\circ$] & 203.8 & 197.3 & 190.8 & 76.3 & 192.3 & [113.5-270.3] & [26.3-337.0] & [1.1-358.6] & $\mathcal{{U}}$\\          \hline
      $a_\mathrm{S}$ & [AU] & 0.000002853 & 0.000002750 & 0.000002756 & 0.000000426 & 0.000002755 & [0.000002329-0.000003180] & [0.000001904-0.000003612] & [0.000001437-0.000004076] & --\\
      $a$ & [AU] & 0.039102 & 0.039416 & 0.039436 & 0.000235 & 0.039435 & [0.039200-0.039668] & [0.038960-0.039906] & [0.038709-0.040128] & --\\
      $m$ & [$M_\oplus$] & 13.26 & 13.03 & 13.03 & 2.02 & 13.02 & [11.01-15.05] & [8.99-17.09] & [6.73-19.17] & --\\
      $m$ & [$M_\mathrm{{J}}$] & 0.04172 & 0.04100 & 0.04100 & 0.00636 & 0.04098 & [0.03464-0.04736] & [0.02828-0.05378] & [0.02116-0.06033] & --\\
      $m$ & [$M_\odot$] & 0.00003982 & 0.00003913 & 0.00003914 & 0.00000607 & 0.00003911 & [0.00003306-0.00004521] & [0.00002699-0.00005133] & [0.00002020-0.00005759] & --\\
      $T_\mathrm{C}$ & [BJD] & 2455501.200 & 2455502.749 & 2455502.399 & 0.918 & 2455502.564 & [2455501.511-2455503.286] & [2455500.131-2455503.723] & [2455500.011-2455503.815] & --\\
      $T_\mathrm{P}$ & [BJD] & 2455500.16 & 2455500.42 & 2455500.00 & 1.09 & 2455500.02 & [2455498.72-2455501.29] & [2455498.18-2455501.83] & [2455498.09-2455501.91] & --\\
      \hline
    \end{tabular}}
  \end{center}
\end{sidewaystable*}

\end{appendix}
\end{document}